\documentclass[runningheads]{llncs}
\usepackage[T1]{fontenc}
\usepackage{graphicx}
\usepackage[hidelinks]{hyperref}

\usepackage[inline]{enumitem}
\usepackage{pifont}
\usepackage[nolist]{acronym} 

\usepackage{nicematrix}

\usepackage[sort]{cite}

\usepackage{xcolor}

\newcommand{\C}[1]{{(\textbf{#1})}}

\usepackage{tikz}
\usetikzlibrary{calc}
\newcommand{\sw}{0.8em}
\newcommand{\squareF}[1]{%
\tikz{%
\fill[black] rectangle(\sw,#1*\sw);
\draw[line width=0.3pt] rectangle(\sw,\sw);%
}}%

\newcommand{\rate}[1]{{%
\ifnum#1=0 \squareF{0}%
	\else\ifnum#1=1 \squareF{0.25}%
	\else\ifnum#1=2 \squareF{0.5}%
	\else\ifnum#1=3 \squareF{0.75}%
	\else\ifnum#1=4 \squareF{1}%
	\fi\fi\fi\fi\fi%
}}

\usepackage{tikz}
\newcommand\copyrighttext{
\footnotesize
\hspace{5pt}

\hrule
\vspace{2pt}
Authors' version of a paper accepted for publication in \emph{Proceedings of the 20th EAI International Conference on Mobile and Ubiquitous Systems: Computing, Networking and Services (MobiQuitous)}. Cite as: J.~Bodenhausen, C.~Sorgatz, T.~Vogt, K.~Grafflage, S.~Rötzel, M.~Rademacher, and M.~Henze, ``Securing Wireless Communication in Critical Infrastructure: Challenges and Opportunities'', in \emph{Proceedings of the 20th EAI International Conference on Mobile and Ubiquitous Systems: Computing, Networking and Services (MobiQuitous)}, 2023.
}

\newcommand\copyrightnotice{%
	\begin{tikzpicture}[remember picture,overlay] \node[anchor=south,yshift=30pt] at (current page.south)
	{{\parbox{\dimexpr\textwidth-\fboxsep-\fboxrule\relax}{\copyrighttext}}};
	\end{tikzpicture}%
}

\renewcommand{\orcidID}[1]{\href{https://orcid.org/#1}{\includegraphics[width=8pt]{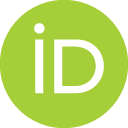}}}

\newcommand{\highlight}[1]{\noindent\textbf{#1}}

\begin{document}
\title{Securing Wireless Communication in Critical Infrastructure: Challenges and Opportunities}
\titlerunning{Securing Wireless Communication in Critical Infrastructure}
\author{%
Jörn Bodenhausen\inst{1}\orcidID{0009-0007-7055-0541} \and
Christian Sorgatz\inst{2}\orcidID{0009-0005-1504-7033} \and
Thomas Vogt\inst{1} \and
Kolja Grafflage \inst{3} \and 
Sebastian Rötzel \inst{4} \and 
Michael Rademacher\inst{2}\orcidID{0000-0002-3721-370X} \and
Martin Henze\inst{1,2}\orcidID{0000-0001-8717-2523}}
\authorrunning{J. Bodenhausen et al.}
\institute{Security and Privacy in Industrial Cooperation, RWTH Aachen University, Aachen, Germany,
\email{\{bodenhausen, vogt, henze\}@spice.rwth-aachen.de}
\and
Cyber Analysis \& Defense, Fraunhofer FKIE, Wachtberg, Germany,
\email{\{christian.sorgatz, michael.rademacher, martin.henze\}@fkie.fraunhofer.de}
\and
ATS Elektronik GmbH, Wunstorf, Germany, \email{kolja.grafflage@atsonline.de}
\and
Bonn-Netz GmbH, Bonn, Germany, \email{sebastian.roetzel@bonn-netz.de}
}
\maketitle              %
\begin{abstract}
Critical infrastructure constitutes the foundation of every society.
While traditionally solely relying on dedicated cable-based communication, this infrastructure rapidly transforms to highly digitized and interconnected systems which increasingly rely on wireless communication.
Besides providing tremendous benefits, especially affording the easy, cheap, and flexible interconnection of a large number of assets spread over larger geographic areas, wireless communication in critical infrastructure also raises unique security challenges.
Most importantly, the shift from dedicated private wired networks to heterogeneous wireless communication over public and shared networks requires significantly more involved security measures.
In this paper, we identify the most relevant challenges resulting from the use of wireless communication in critical infrastructure and use those to identify a comprehensive set of promising opportunities to preserve the high security standards of critical infrastructure even when switching from wired to wireless communication.

\end{abstract}

\copyrightnotice

\vspace{-2.7em}

\section{Introduction}

Critical infrastructure encompasses all physical and cyber assets that are essential to maintain vital societal functions. Common examples are smart grids (i.e., electricity and water distribution)~\cite{mecheva2020cybersecurity,tuptuk2021systematic} and smart city services (i.e., intelligent traffic and transportation)~\cite{7823349,hamid2019cyber}.
Over the past decades, this infrastructure has transformed towards highly digitized and interconnected intelligent systems \cite{vandervelde2020medit,eggert2014sensorcloud,wolsing2022ipal} to capitalize on benefits such as efficiency improvements, better visibility into the network, and increased flexibility \cite{bader2023metrics}, while at the same time meet increasing complexity, e.g., resulting from distributed energy resources~\cite{krause2021cybersecurity,bader2023wattson}.

Traditionally, critical infrastructure relied on private wired communication infrastructure to interconnect assets (e.g., fiber optic networks)~\cite{krause2021cybersecurity}.
More recently, we observe an increasing incorporation of wireless communication in critical infrastructure, supplementing and partially replacing wired communication, fueled by various complementing developments \cite{henze2017network}:
\begin{enumerate*}[label=\protect\ding{\value*},start=172]
\item Significant advances in the fundamental technology underlying wireless communication enable increased throughput and low latency~\cite{rappaport2017overview,luo2020empirical}, energy efficient transmissions over long-distances~\cite{rademacher2021path}, and a significantly higher density of devices~\cite{9032817}.
\item A fundamental shift to highly digitized and interconnected systems, where many existing facilities are not connected to a private wired network yet and new emerging entities that do not have an established dedicated wired network at all~\cite{krause2021cybersecurity}.
\item The emergence of new usage scenarios where wired communication is technically or economically infeasible, e.g., in the context of smart cities and \ac{IoT} deployments, where large numbers of mobile and resource-constrained devices are deployed over large areas~\cite{serror2020challenges}.
\end{enumerate*}

While the resulting increasing use of \emph{wireless} communication in critical infrastructure provides tremendous benefits, it also poses serious challenges for security.
Most importantly, critical communication might now take place over publicly accessible or at least visible networks outside the physical control of the critical infrastructure operator, much unlike the traditional security assumptions of dedicated private networks with strong perimeter protection~\cite{krause2021cybersecurity}.
Consequently, to sustainably capitalize on the advantages of wireless communication in critical infrastructure, a thorough understanding of these resulting security challenges as well as opportunities to address them is required.

\highlight{Related Work.}
Various works study and systematize the security challenges resulting from modern wireless communication (not specific to critical infrastructure)~\cite{rathore2017review,somasundaram2021review,touqeer2021smart}, (mostly) wired communication in distinct critical infrastructure~\cite{alcaraz2015critical} such as water systems~\cite{tuptuk2021systematic} or power grids~\cite{vandervelde2020medit,krause2021cybersecurity}, as well as highly digitized and interconnected systems such as smart cities~\cite{hamid2019cyber,mecheva2020cybersecurity} and the (industrial) \ac{IoT}~\cite{serror2020challenges,dahlmanns2022tls}.
While these works provide a comprehensive overview of the respective research fields, they do not generalize the unique challenges of securing wireless communication in critical infrastructure.
Contrary, different works focus on isolated challenges and possible solutions to address these, e.g., by analyzing path-loss of electromagnetic waves in urban environments~\cite{rademacher2021path}, studying the performance of \ac{TLS} for \ac{IoT}~\cite{rademacher2022bounds,restuccia2020low}, evaluating low latency but secure communication for the \ac{IoT}~\cite{hiller2018secure}, and securing \ac{IoT} devices with secure elements~\cite{schlapfer2019security}.
Still, these works do not focus on deriving a comprehensive picture of the challenge and solution space. 
This contrast results in a gap between an overview of relevant challenges for the particular case of wireless communication in critical infrastructure and the potential benefit from solutions for individual challenges. 

\highlight{Contributions.}
In this paper, we fill this gap by comprehensively studying the relevant challenges resulting from the use of wireless communication in critical infrastructure and identifying opportunities to uphold a strong level of security even under these challenging conditions.
More precisely, our contributions are:
\begin{enumerate}[nosep]
	\item We systematize the use of wireless communication in critical infrastructure and provide an overview of relevant technology and its properties (Section~\ref{sec:background}).
	\item We categorize and summarize security challenges that are specifically applicable to wireless communication in critical infrastructure (Section~\ref{sec:challenges}).
	\item We identify and discuss complementary opportunities to address the challenges of secure wireless communication in critical infrastructure (Section~\ref{sec:solutions}).
\end{enumerate}

\section{Wireless Communication in Critical Infrastructure} \label{sec:background}

In recent years, critical infrastructure increasingly relies on wireless communication~\cite{alcaraz2015critical}.
To lay the foundation to identify and address resulting security challenges, we first identify developments and motivations that fuel this trend (cf.\ Section~\ref{sec:background:motivation}), before we introduce and discuss various wireless communication technology that is especially relevant for critical infrastructure (cf.\ Section~\ref{sec:wirelessNetworks}).

\subsection{From Wired to Wireless Communication} \label{sec:background:motivation}

\begin{figure}[t]
\centering
\includegraphics{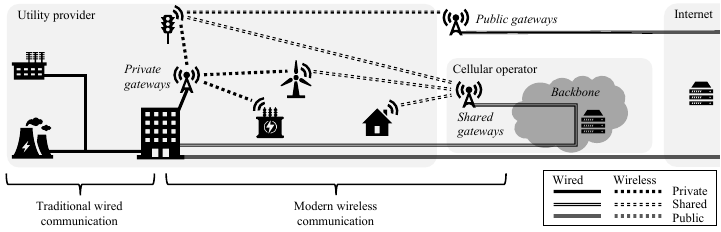}
\caption{Communication in critical infrastructure shifts from dedicated private wired networks to heterogeneous wireless communication, encompassing all means of private, public, and shared communication networks as well as dual-homed devices.} %
\label{fig:overview}
\end{figure}
For decades, critical infrastructure relied on private \emph{wired} communication networks~\cite{krause2021cybersecurity}, as exemplified for a typical (municipal) utility provider in the left part of Figure~\ref{fig:overview}.
With advancements in wireless communication technology and the push towards digitization, critical infrastructure increasingly relies on a confluence of heterogeneous \emph{wireless} communication networks, which are no longer necessarily private, but potentially shared with other critical infrastructure or the public (right part of Figure~\ref{fig:overview}).
While wireless communication will likely serve as a complement to existing wired communication, it mainly serves as an enabler for novel use cases and a remedy for the increasing need for communication capabilities of assets.
Notably, the choice of a wireless communication technology within one critical infrastructure will neither be homogeneous nor exclusive, i.e., multiple technologies will coexist and individual devices or assets might connect over different wireless networks~\cite{7823349}.

Exemplarily focusing on a (municipal) utility provider as depicted in Figure~\ref{fig:overview}, we discuss and highlight the reason and motivation as well as different incarnations of the shift to wireless communication. 
Most notably, the shift towards an increasing use of wireless communication would not be possible without tremendous advancements in the underlying technologies~\cite{rappaport2017overview,9032817}.
These advancements have improved wireless networks, enhancing reliability, efficiency, and capacity, making their integration into critical infrastructure more feasible.

Complementing these developments, there has been a fundamental transformation towards highly digitized and interconnected systems, exemplified by the emergence of smart grids~\cite{dileep2020survey}, where existing facilities that were previously devoid of communication requirements now need connectivity.
Moreover, an increasing amount of entities such as heat pumps, tiny solar power installations, and charging stations have arisen that lack an established dedicated wired network altogether.
This trend is carried to the extreme by completely new smart city scenarios (i.e., smart recycling and parking, intelligent transportation~\cite{7823349}) where wired communication may prove technically or economically unviable. 

Overall, the shift from wired to wireless communication enables a multitude of new use cases and applications.
To illustrate this, we will briefly reference the most prominent applications for secure wireless communication in critical infrastructure. 
In general, applications can be grouped into categories such as energy, environment, industry, living, and services \cite{7823349}.
Here, critical infrastructure mainly affects energy~\cite{dileep2020survey, 7964673}, industry~\cite{8869323}, and services~\cite{7964673}.
The dominant critical infrastructure application is smart grids~\cite{dileep2020survey, 7964673}, i.e., the monitoring of energy generation, transmission, and consumption~\cite{krause2021cybersecurity,vandervelde2020medit,klaer2020graphbased}.
In the field of services, applications arise due to the possibility of mobile sensors, which allows, e.g., for the monitoring of transportation services~\cite{7964673}.
Moreover, the paradigm of Industry 4.0 leads to interesting use cases and demands for very dense wireless networks to interconnect machines and computational capabilities~\cite{glebke2019case,8869323}. 

In summary, the incorporation of wireless communication within critical infrastructure has gained traction due to significant technological advancements, the shift towards interconnected systems such as smart grids, and the rise of new scenarios where wired communication is infeasible.
As depicted in Figure~\ref{fig:overview}, in contrast to wired communication which predominantly relies on dedicated private fiber optic and copper wire networks, wireless communication in critical infrastructure can rely on a plethora of cellular radio technology in private, public, and shared deployments.

\subsection{Heterogeneous Wireless Networks} \label{sec:wirelessNetworks}
The range of different cellular radio technologies potentially suitable for critical infrastructure is immense.
The most common technologies are \ac{LTE}, \ac{NB-IoT}, \ac{GSM}, \ac{TETRA}, \ac{LoRaWAN}, Wi-Fi, 450 MHz \ac{LTE-M}, 5G and in the future 6G.
Choosing a suitable technology is challenging since they have widely differing characteristics and capabilities. 

What all these wireless networks have in common is, that they are cellular radio networks, i.e., their topology can be summarized as interconnected base stations.
Each base station typically serves a certain area, while overlaps are possible and desired.
Base stations require an exposed location, a stable power supply, and a backhaul connection typically realized via wired networks.

The characteristics and capabilities of the different technologies are mainly determined by the maturity level of the technology, the frequency used, and the available bandwidth.
At Sub-GHz bands (e.g., 450 MHz LTE-M or \ac{LoRaWAN}), a higher range (cell size) and better penetration of structures can be achieved.
However, the bandwidth is scarce which reduces throughput and scalability and therefore limits the potential applications~\cite{falanji2022range,rademacher2022bounds,thomassen2022study}.
At higher frequency bands (e.g., mmWave for 5G), the throughput is comparable or even superior to wired technologies but the range of each base station and device is limited~\cite{rappaport2017overview}.

In wireless communication, frequency bands are heavily regulated.
There are frequency bands that are strictly assigned to certain entities, i.e., \acp{MNO}, while others are generally accessible.
Receiving the exclusive right to use a certain band (e.g., 450 MHz LTE-M, 5G) requires significant investment costs, which need to be generated.
However, due the exclusive control, the operator can also guarantee a certain \ac{QoS}.
License-exempt bands reduce the operation costs (e.g., Wi-Fi, \ac{LoRaWAN}).
However, particularly in urban scenarios, these bands can become uncontrollably crowded which compromises their applicability for critical infrastructure~\cite{rademacher2018quantifying}.

Similar considerations hold for the operation model of the different wireless technologies.
As visualized in Figure~\ref{fig:overview}, there are three possibilities: public, shared, and private. 
Interestingly, there are technology advancements such as 5G network slicing which blur the boundaries between public and private network infrastructures~\cite{9389979}.
\emph{Public} network infrastructure is operated by entities (\acp{MNO}) not directly associated with critical infrastructure.
\ac{3GPP} technologies (\ac{LTE}, 5G) are widely used and the network is shared with others (e.g., smartphones).
Recently, a new type of public network infrastructure, driven by a shared community approach, emerged.
Most prominently, The Things Network~\cite{TheThing20:online} builds a wide-range \ac{LoRaWAN} network.
\emph{Shared} infrastructure is operated by entities directly associated to critical infrastructure.
While building and operating a wireless network is not feasible for a single critical infrastructure operator, multiple operators can form an association and share costs, e.g., the nationwide 450 MHz \ac{LTE} network for critical infrastructure in Germany~\cite{BnetzA450MHz}.
Compared to public networks, the network is not shared outside the association.
\emph{Private} networks are directly operated by critical infrastructure operators.
These networks are limited to certain areas where a huge device density renders a network economically feasible (e.g., plants, smart cities) or where there is a lack of other solutions (e.g., offshore wind farms).
Technologies operating in license-exempt bands are preferred (e.g., Wi-Fi, \ac{LoRaWAN}), however, private 5G campus networks provide an interesting alternative.

Overall, it can be stated that due to their versatility, heterogeneous wireless networks offer enormous potential to be utilized for a wide variety of applications. 
However, this versatility also leads to significant complexity that affects operators and their clients.
The huge diversity in wireless technologies, frequency regulation, and operation models not only complicates the deployment and operation of wireless communication in critical infrastructure but also fundamentally challenges security.

\section{Challenges of Securing Wireless Communication} \label{sec:challenges}
Wireless communication in critical infrastructure places high demands on security.
At the same time, various significant challenges must be addressed that differ depending on the particular setting and influences such as operation models, particular technologies, network topology, and utilized devices. %
Thus, each particular use case can shift the focus to certain challenges or create new ones. 

\highlight{Methodology}. We strive to identify those challenges that are particularly relevant to wireless communication in critical infrastructure.
To this end, we study various literature providing an overview of security challenges resulting from various aspects relevant to wireless communication in critical infrastructure:
\begin{enumerate*}[label=(\roman*)]
\item modern wireless communication in general, i.e, not focusing on critical infrastructure~\cite{rathore2017review,somasundaram2021review,touqeer2021smart}
\item wired communication in specific, isolated branches of critical infrastructure~\cite{alcaraz2015critical,tuptuk2021systematic,vandervelde2020medit,krause2021cybersecurity}, as well as
\item fully interconnected and digitized intelligent (future) systems~\cite{hamid2019cyber,mecheva2020cybersecurity,serror2020challenges,dahlmanns2022tls,dahlmanns2020easing}.
\end{enumerate*}
We complement these more holistic surveys with highly specialized works focusing on isolated challenges of securing wireless communication in critical infrastructure, especially with respect to
\begin{enumerate*}[label=(\roman*),start=4]
\item network and device constraints alongside various dimensions~\cite{rademacher2021path,elayoubi2008performance,rademacher2022bounds,restuccia2020low},
\item a wide range of security challenges~\cite{mecheva2020cybersecurity,osanaiye2018denial,echeverria2019authentication,rfcTLSRecommendations,roy2023device}, as well as
\item legal and deployment considerations~\cite{BSI-Messsysteme,alcaraz2015critical}.
\end{enumerate*}

Overall, using this methodology, we identify twelve overarching challenges of securing wireless communication in critical infrastructure, which we broadly categorize into four groups:
\begin{enumerate*}[label=\protect\ding{\value*},start=172]
\item constraints of the utilized network and devices (Section \ref{sec:constraints}),
\item security of networks (Section \ref{sec:networkSecurityConcerns}),
\item security of the utilized devices (Section \ref{sec:deviceSec}), and
\item application-specific challenges (Section \ref{sec:applicationSpecific}). 
\end{enumerate*}
To provide the foundation to adequately address these challenges, in the following, we further discuss and analyze them in more detail.

We intentionally chose this ordering and categorization to make the section structured and coherent. 
Our ordering does not necessarily resemble the difficulty or importance of the identified challenges, which mostly depends on the specific situation and perspective. 
Thus, some of those challenges are more fundamental and tangible while others might be considered rather conceptual. 
For instance, the challenges grouped in the categories \protect\ding{172} and \protect\ding{174} could be considered fundamental, as these stem from the devices as well as the environment and are thus entirely rigid. 
The challenges in the categories \protect\ding{173} and \protect\ding{175} on the other hand are not as rigorous and depend more on design decisions and requirements than on actual limitations. 
In any case, a detailed use case analysis of each particular application and its environment is necessary to determine the magnitude and importance of each challenge and thus derive a specific prioritization.

\subsection{Network and Device Constraints} \label{sec:constraints}

Securing wireless communication is first and foremost challenged by constraints of the employed network and devices.
Most importantly, wireless networks are challenged w.r.t. reliability and mobility.
Furthermore, especially in \ac{IoT} settings, both network and devices can be restricted with regard to available resources such as bandwidth, computing power, or energy.

\highlight{C1 - Reliability} is a crucial aspect for critical infrastructure, where seamless communication and real-time data transmission are paramount %
to prevent disruptions and ensure smooth operation and safety.
Any interruptions or failures in the wireless network can have far-reaching consequences, including power outages, equipment malfunctions, and delays in emergency response.
Therefore, building and maintaining a reliable mobile network infrastructure is essential to ensure the smooth operation of critical infrastructure systems. %
Due to the inherent characteristics of wireless communication, this is particularly difficult to achieve as wireless connections are susceptible to various environmental factors that can drastically affect signal strength and quality.
Most importantly, obstacles such as terrain, buildings, or other structures as well as distance can reduce the reliability of the wireless connection by causing signal attenuation and propagation loss~\cite{rademacher2021path}.
Furthermore, wireless networks are prone to interference caused by other radio technologies operating on the same frequencies or from inter-cell interference caused by the same technology~\cite{elayoubi2008performance}. %
These difficulties must be considered during the planning of any mobile network intended for use in critical infrastructure.

\highlight{C2 - Mobility}
additionally leads to non-uniform connection quality that depends on the device position and corresponding local influences on reliability, influenced by factors such as distance from the base station, physical obstructions, device interference, and signal propagation characteristics~\cite{rademacher2021path}.
This variability in network quality poses challenges for both mobile nodes and devices that are permanently deployed in unfavorable positions.
Seamless roaming between different coverage areas is critical to maintaining an uninterrupted connection when devices move between cells.
A challenge for devices in motion is the potentially increased power consumption when they remain connected to a cell at its edge~\cite{tayyab2019survey}, despite a better signal being available from another nearby cell~\cite{kanwal2015reduced}.
The ability to switch between cells presents an opportunity for attackers to target \acp{UE} by using fake/rogue base stations~\cite{hussain2018lteinspector}.
Another challenge with mobility is the availability of wireless technology in any location where the device may be present.
Therefore, not only roaming between cells but also between wireless technologies is an important aspect to address.
The security vulnerabilities of all supported technologies become critical when devices have the ability to switch between them, especially if they can be forced into switching from external sources.
 For instance, if 5G coverage is limited, then a device or \ac{UE} can downgrade to an earlier generation of mobile communication, such as \ac{LTE} or \ac{GSM}, so that the device can maintain uninterrupted access to the network.

\highlight{C3 - Network Limitations} imposed by the particular technology can further restrict the network beyond the challenges of wireless technologies in general.
Most notably, the throughput is limited compared to a wired network since throughput requires bandwidth and spectrum is scarce.
This limitation holds for licensed (e.g., 450 MHz LTE~\cite{SmartGridCom}) as well as unlicensed bands (e.g., LoRaWAN~\cite{rademacher2022bounds}) and in particular restricts the acceptable bandwidth overhead that can be added by security mechanisms. 
Furthermore, devices compete for this bandwidth, which becomes a problem when particularly many devices are part of the network, as the available bandwidth is shared among all devices and thus even more limited.
This challenge was exemplarily illustrated for \ac{LoRaWAN}, where numerous devices might even make a timely \ac{TLS} connection establishment infeasible \cite{rademacher2022bounds}.
As a result, latency is also an issue, as it takes longer to transmit the full message, which can have significant implications for safety-related mechanisms such as an emergency shutdown.
This is reinforced by the technology-specific maximum message size (e.g., 256 bytes for \ac{LoRaWAN}), through which larger messages might be fragmented and delayed even if the wireless network is currently not busy.
Furthermore, the communication range of different wireless technologies can differ significantly and thus restrict their usability (cf. Section \ref{sec:wirelessNetworks}).

\highlight{C4 - Device Limitations} further impose challenges beyond the limitations of the network.
Especially in an \ac{IoT}-like setting, device resources such as compute power, memory, storage, or energy can be rather restricted \cite{rfcConstrained}.
Such limitations can have a significant impact on the system.
For instance, they can limit the applicability of authentication and encryption mechanisms, especially asymmetric ones, as these are commonly computationally expensive \cite{restuccia2020low}.
Memory constraints limit the possible complexity of a system and its ability to cache information. 
Moreover, especially for battery-powered devices, energy efficiency is of utmost importance.
As transmission and reception are particularly energy-intensive, messages should be kept as short as possible and bandwidth efficiency is crucial. 
All those factors also limit the applicability of mechanisms that address security concerns.

\subsection{Network Security Concerns} \label{sec:networkSecurityConcerns}

While the challenges discussed so far primarily affect the usability of wireless technology, concerns also arise for the security of transmitted data.
Naturally, conventional security considerations, i.e., the CIA triad consisting of confidentiality, integrity, and availability, also apply here.
Moreover, proper authentication is a fundamental challenge that needs to be addressed. 

\highlight{S1 - Confidentiality and Integrity} protect data from unauthorized access (confidentiality) and unauthorized alteration (integrity).
Such protection is particularly relevant for wireless communication in critical infrastructure, where sensitive data from both the customer and the critical infrastructure operator is transmitted over a wireless medium that can be accessed by anyone in communication range. 
In some cases, such as the transmission of generic machine instructions, confidentiality might be optional, but integrity is of utmost importance to mitigate the manipulation and insertion of messages. 
However, assuring confidentiality and integrity can be challenging in constrained environments, as suitable security mechanisms add overhead~\cite{mecheva2020cybersecurity}.
Furthermore, it must be considered that hop-by-hop mechanisms, as usually integrated into wireless technology, do not suffice if the subsequently traversed infrastructure is not private, often calling for the use of end-to-end security. 

\highlight{S2 - Availability} ensures that resources can be accessed when needed. 
Unlike wired networks, the wireless medium can be used easily and unrestricted by anyone in range, allowing for a multitude of attacks on availability.
Malicious interference in the form of jamming can cause a disturbance in the wireless network such that legitimate communication is partially or completely disrupted~\cite{mpitziopoulos2009survey}.
Moreover, energy depletion of battery-powered devices is possible by preventing the system from sleeping~\cite{osanaiye2018denial}. 
Hence, protecting wireless networks from such attacks is exceptionally challenging, especially considering critical infrastructure.

\highlight{S3 - Authentication} \label{sec:authentication} ensures that only authorized devices participate in communication. %
While wired networks can be secured to some extent through their architecture and segmentation, cryptographic methods must inevitably be used in a wireless network to ensure that only authenticated devices can access it.
Such cryptographic methods usually rely on asymmetric mechanisms, which are demanding for resource-constrained devices~\cite{krause2021cybersecurity}.
Furthermore, proper authentication is required in various contexts (e.g., network connection, end-to-end security, system updates ~\cite{henze2017distributed}). 
Overall, such methods must therefore be designed in a resource-efficient way and devices must receive appropriate credentials.

\subsection{Device Security and Challenges} \label{sec:deviceSec}

The changes in security requirements and network environment due to wireless communication in critical infrastructure also impose challenges on individual devices.
One such challenge is the more difficult deployment with cryptographic material. %
As credentials weaken over time or become otherwise insecure, revocation mechanisms must be utilized.
Moreover, easier physical access of adversaries to devices deployed in more or less public spaces must be considered.

\highlight{A1 - Deployability} of devices is significantly impeded by the need for credentials for subsequent authentication.
In particular root credentials are required to authenticate all subsidiary credentials and, moreover, all connections and firmware upgrades. %
Devices need to be equipped and configured with root credentials during production or deployment \cite{echeverria2019authentication}. %
This deployment may be automated or require manual work, but in any case, makes the deployment process more difficult and expensive.

\highlight{A2 - Device Exclusion} allows to remove previously authorized devices from a network, e.g., when credentials get stolen or otherwise compromised as well as when underlying cryptographic primitives lose their promised security level. 
Especially credential theft could occur rather frequently in critical infrastructure due to the easier physical access to devices that are increasingly deployed in public spaces. 
For such cases, there is a need for revocation mechanisms and by extension the exclusion of compromised devices from the network.
While there are solutions that address this issue for subsidiary credentials, they usually require trust in the utilized root credentials. 
Protection and revocation of root credentials thus is an open challenge \cite{rfcTLSRecommendations,henze2014trust,henze2013maintaining}. 

\highlight{A3 - Physical Access to Devices} refers to an adversary's ability to physically extract, e.g., keying material from a device.
While devices were conventionally inaccessible on private premises, they are increasingly being deployed in openly accessible spaces \cite{roy2023device}, e.g., in the context of smart cities.
An illustrative example would be smart metering systems, that are deployed in private homes and are thus easily accessible to the residents. 
Furthermore, the wireless characteristic allows for devices to be moved while still being connected to the network. 
This results in special requirements and challenges for the physical protection of the devices themselves and the critical information stored on them. 

\subsection{Application-Specific Requirements} \label{sec:applicationSpecific}

In addition to these challenges that follow from the use of wireless communication in critical infrastructure in general, further use-case-specific requirements might need consideration. 
Most importantly, this concerns potential legal requirements as well as a certain flexibility to allow for seamless operation. 

\acused{BSI}

\highlight{R1 - Legal Requirements} are imposed on a wide range of domains and in particular critical infrastructure. 
For example, a guideline of the German Federal Office for Information Security (\ac{BSI}) that is applicable for smart meter infrastructure \cite{BSI-Messsysteme} demands the use of \ac{TLS} with mutual certificate-based authentication, limits session lifetime to 48 hours, and restricts the choice of cryptographic algorithms and parameters.  
Such requirements can also cover aspects outside security, e.g., functional safety or interoperability, and significantly limit the possibility of specialized and optimized solutions. 

\highlight{R2 - Flexibility} ensures the seamless operation of the system. 
For once, the interoperability of the system must be considered, since the infrastructure might not be controlled by a single entity. %
Moreover, the updatability of and access to the system, i.e., serviceability, must be taken into account~\cite{alcaraz2015critical}.
As it might be impracticable or impossible to access devices that are deployed in the field, remote access and update solutions must be utilized.
Furthermore, the use of several wireless technologies in parallel could enhance the redundancy and robustness of the system. 
Lastly, as the use case might not yet be precisely defined, the system should allow for adaptability to changing conditions.

\section{Opportunities to Secure Wireless Communication} \label{sec:solutions}

\begin{table}[t]
	\centering
	\caption{Various opportunities to secure wireless communication in critical infrastructure exist to address the identified fundamental security challenges (marked by \rule{0.8em}{0.8em}).}
	\label{tab:mapping}
	\begin{NiceTabular}{l|lllll|llll|llll|lll|}
	\diagbox{\textbf{Opportunities}\\to secure wireless\\communication in critical\\infrastructure (Section \ref{sec:solutions})}{\textbf{Challenges} of securing\\ wireless communication in \\critical infrastructure\\ (Section \ref{sec:challenges}) }
									   & \rotatebox{90}{\textbf{Netw./Dev. Constr.}} & \rotatebox{90}{C1 - Reliability}  & \rotatebox{90}{C2 - Mobility}  & \rotatebox{90}{C3 - Network Limit.} & \rotatebox{90}{C4 - Device Limit.}  & \rotatebox{90}{\textbf{Netw. Sec. Concerns}} & \rotatebox{90}{S1 - Conf./Integrity}  & \rotatebox{90}{S2 - Availability}  & \rotatebox{90}{S3 - Authentication}  &\rotatebox{90}{\textbf{Dev. Security/Chall.}}&\rotatebox{90}{A1 - Deployability}  & \rotatebox{90}{A2 - Device Exclusion}  & \rotatebox{90}{A3 - Phy. Acc. to Dev.}  & \rotatebox{90}{\textbf{App.-Spec. Require.}} & \rotatebox{90}{R1 - Legal Require.}  & \rotatebox{90}{R2 - Flexibility}  \\ \hline
	\textbf{Network Security}          &                                             &                                   &                                &                                     &                                     &                                              &                                       &                                    &                                      &                                            &                                     &                                        &                                         &                                              &                                      &                                   \\
	Lower Layer Security Mechanisms    &                                             & \rate{4}                          & \rate{0}                       & \rate{4}                            & \rate{0}                            &                                              & \rate{0}                              & \rate{0}                           & \rate{0}                             &                                            & \rate{0}                            & \rate{0}                               & \rate{0}                                &                                              & \rate{0}                             & \rate{0}                          \\
	Secure Communication Channel       &                                             & \rate{0}                          & \rate{0}                       & \rate{4}                            & \rate{4}                            &                                              & \rate{4}                              & \rate{0}                           & \rate{0}                             &                                            & \rate{0}                            & \rate{0}                               & \rate{0}                                &                                              & \rate{0}                             & \rate{0}                          \\ \hline
	\textbf{Communication Security}    &                                             &                                   &                                &                                     &                                     &                                              &                                       &                                    &                                      &                                            &                                     &                                        &                                         &                                              &                                      &                                   \\
	Resource-Conscious E2E Security    &                                             & \rate{0}                          & \rate{0}                       & \rate{4}                            & \rate{4}                            &                                              & \rate{4}                              & \rate{0}                           & \rate{0}                             &                                            & \rate{0}                            & \rate{0}                               & \rate{0}                                &                                              & \rate{4}                             & \rate{4}                          \\
	E2E Security Extensions            &                                             & \rate{0}                          & \rate{0}                       & \rate{4}                            & \rate{4}                            &                                              & \rate{0}                              & \rate{0}                           & \rate{0}                             &                                            & \rate{0}                            & \rate{0}                               & \rate{0}                                &                                              & \rate{0}                             & \rate{0}                          \\
	Cryptography Optimizations         &                                             & \rate{0}                          & \rate{0}                       & \rate{4}                            & \rate{4}                            &                                              & \rate{4}                              & \rate{0}                           & \rate{0}                             &                                            & \rate{0}                            & \rate{0}                               & \rate{0}                                &                                              & \rate{0}                             & \rate{0}                          \\ \hline
	\textbf{Device Security}           &                                             &                                   &                                &                                     &                                     &                                              &                                       &                                    &                                      &                                            &                                     &                                        &                                         &                                              &                                      &                                   \\
	Hardware Security Mechanisms       &                                             & \rate{0}                          & \rate{0}                       & \rate{0}                            & \rate{4}                            &                                              & \rate{0}                              & \rate{0}                           & \rate{0}                             &                                            & \rate{0}                            & \rate{0}                               & \rate{4}                                &                                              & \rate{4}                             & \rate{0}                          \\ 
	Authenticated Code and Secure Boot &                                             & \rate{0}                          & \rate{0}                       & \rate{0}                            & \rate{0}                            &                                              & \rate{0}                              & \rate{0}                           & \rate{4}                             &                                            & \rate{0}                            & \rate{0}                               & \rate{4}                                &                                              & \rate{0}                             & \rate{0}                          \\
	Root of Trust                      &                                             & \rate{0}                          & \rate{0}                       & \rate{0}                            & \rate{0}                            &                                              & \rate{0}                              & \rate{0}                           & \rate{4}                             &                                            & \rate{4}                            & \rate{4}                               & \rate{4}                                &                                              & \rate{0}                             & \rate{4}                          \\ \hline
	\textbf{Network Knowledge}         &                                             &                                   &                                &                                     &                                     &                                              &                                       &                                    &                                      &                                            &                                     &                                        &                                         &                                              &                                      &                                   \\
	Propagation Conditions             &                                             & \rate{0}                          & \rate{4}                       & \rate{4}                            & \rate{0}                            &                                              & \rate{0}                              & \rate{4}                           & \rate{0}                             &                                            & \rate{0}                            & \rate{0}                               & \rate{0}                                &                                              & \rate{0}                             & \rate{0}                          \\
	Context Awareness                  &                                             & \rate{0}                          & \rate{4}                       & \rate{4}                            & \rate{4}                            &                                              & \rate{0}                              & \rate{0}                           & \rate{0}                             &                                            & \rate{0}                            & \rate{0}                               & \rate{4}                                &                                              & \rate{0}                             & \rate{0}                          \\
	Roaming and Redundancy             &                                             & \rate{4}                          & \rate{4}                       & \rate{0}                            & \rate{0}                            &                                              & \rate{4}                              & \rate{4}                           & \rate{0}                             &                                            & \rate{0}                            & \rate{0}                               & \rate{0}                                &                                              & \rate{0}                             & \rate{0}                          \\ \hline
	\end{NiceTabular}
	\end{table}

To address the identified crucial challenges for securing wireless communication in critical infrastructure, we move forward by identifying and examining potential opportunities to secure wireless communication but also point out their shortcomings and promising potential for further research.
For easier reference, Table \ref{tab:mapping} gives an overview of the challenges that are addressed by the various discussed approaches, which we structured into four categories:
\begin{enumerate*}[label=\protect\ding{\value*},start=172]
\item security for the network technology (Section \ref{sec:networkSecurity}),
\item security for the encapsulated communication (Section \ref{sec:communicationSecurity}),
\item security of the devices themselves (Section \ref{sec:deviceSecurity}), and
\item enhancing security using detailed knowledge of the network (Section \ref{sec:networkKnowledge}).
\end{enumerate*}

\subsection{Network Security} \label{sec:networkSecurity}

Wireless networks are built upon diverse technologies, each equipped with its own set of inherent security mechanisms.
As critical infrastructure becomes more interconnected, wireless network security becomes increasingly important.

\highlight{Lower Layer Security Mechanisms} are particularly relevant for wireless communication, i.e., using a shared medium.
 Even if end-to-end security on a higher layer is utilized, the lower layer security of the particular communication technology has to be addressed, for which various mechanisms exist.
With \ac{LTE} and 5G, etc., radio traffic typically takes place in licensed frequencies and heavy regulations ensure reduced interference from other radio technologies. 
To further mitigate unintended interference, frequency planning~\cite{elayoubi2008performance} can be used. 
To combat jamming, employing advanced modulation schemes is an effective countermeasure~\cite{jaitly2017security}.
Thus, reliability \C{C1} can be improved~\cite{alcaraz2015critical} while addressing network limitations \C{C3}.

\highlight{Secure Communication Channels} ensure that data is protected while traversing a network.
To secure data transport from the \ac{MNO} to the customer, a \ac{VPN} or a dedicated physical connection can be used.
This can be combined with private \acp{APN} or network slicing~\cite{9389979} on the access part to isolate data from different classes of users~\cite{GSMA-Features}.
\ac{IoT} devices typically only require communication with specific servers, allowing to restrict communication, which mitigates the risks of integrating compromised \acp{UE} into botnets as it would require communication with unauthorized servers.
Such mechanisms address confidentiality and integrity \C{S1} on the network layer by establishing secure channels for data transmission, ensuring that data remains confidential and intact during network traversal.
However, on this layer, it is particularly important to address network and device limitations \C{C3, C4}.
\ac{IoT} devices often send little data, making it inefficient to send it as an \ac{IP} packet.
Cellular mobile technologies such as \ac{LTE-M} or \ac{NB-IoT} provide \ac{NIDD} as a security mechanism, where data packets are sent without \ac{IP} address and other headers~\cite{GSMA-Features}, eliminating typical \ac{IP}-based attack scenarios.
However, for \ac{NIDD} and similar approaches, existing implementations for higher-layer protocols need to be adapted to work with changes to the layer directly below \cite{8611848}.

\subsection{Communication Security} \label{sec:communicationSecurity}

Security mechanisms and authentication provided by utilized communication technology commonly operate in a hop-by-hop fashion.
However, especially for public networks, further protection mechanisms are required to secure the communication between nodes in an end-to-end manner \cite{dahlmanns2021transparent}.
The commonly used approach to realize this is \ac{TLS} \cite{rfcTLS13}, which is a secure and application-independent protocol that provides confidentiality, integrity protection, and moreover authentication, if a proper root of trust is used for both peers, for end-to-end communication.
However, such approaches cause additional overhead, especially with respect to bandwidth and computing power.

\highlight{Resource-Conscious End-to-End Security} aims to provide confidentiality \C{S1} while respecting limitations of devices \C{C4} and the network \C{C3}. 
In critical infrastructure, \ac{TLS} is often without alternatives due to legal requirements \C{R1} and the requirement for interoperability with other systems \C{R2}.
However, the overhead of \ac{TLS} is not fixed and varies depending on the configuration and utilized cryptographic algorithms \cite{rademacher2022bounds,restuccia2020low}, allowing for optimization through proper configuration and selection of parameters~\cite{rademacher2022bounds}.
Still, further research is required to determine the most beneficial options in a certain scenario and verify the effectiveness of resulting optimizations. 
In this context, it has to be considered that some options for optimization come at the cost of reducing the interoperability and updatability.

\highlight{End-to-End Security Extensions} provide further optimizations beyond what is possible through parametrization and algorithm choice alone.
For instance, the optional \ac{TLS} session resumption mechanism makes subsequent connections more efficient by cryptographically linking them to the initial connection \cite{rfcTLS13}.
Session sharing between applications allows to make better use of session resumption, conserve system resources \C{C4}, and improve latency \cite{hiller2019case}.
Likewise, latency can be improved by pre-computing certain cryptographic operations \cite{hiller2018secure}.
Lastly, caching of information allows to reduce the amount of transmitted data, further countering network limitations \C{C3}.
For \ac{TLS}, e.g., an extension allows to cache parts of the handshake, which can then be omitted in subsequent handshakes \cite{rfcCachedInfo}.
Still, further research is required to assess the practical application and interoperability of these mechanisms, in particular for critical infrastructure depending on the device and the network properties like bandwidth.
This is required as especially interoperability is low for multiple of the extensions proposed for constrained devices.

\highlight{Cryptography Optimizations} allow to further tailor security when communicating only with a limited set of services (at the cost of flexibility). 
However, such solutions must be adequate for and tailored to the particular setting and requirements. %
For instance, if confidentiality is not required, a solution focused solely on integrity protection \C{S1} could significantly reduce resource requirements.
To this end, aggregation of message authentication codes \cite{wagner2024aggregate}, especially using progressive message authentication \cite{armknecht2020promac,wagner2022promac}, improves the bandwidth overhead \C{C3} of integrity protection \C{S1} through cleverly designed shorter integrity tags, where the reduced security of shorter tags is improved upon over time.
BP-MAC \cite{wagner2022bpmac}, on the other hand, precomputes integrity protection tags to better utilize computing resources \C{C4} and significantly improve latency.
From a different perspective, RePeL \cite{wagner2023repel} leverages unused header fields of (industrial) communication protocols to embed integrity protection tags to retrofit integrity protection \C{S1} in the face of network limitations \C{C3}.
However, such approaches generally do not allow for communication with arbitrary systems, as is the case for standardized solutions such as \ac{TLS}.
To enable severely resource-constrained devices to still communicate securely using \ac{TLS}, delegation approaches offload the particularly resource-intensive connection establishment to a more powerful, trusted device \cite{hummen2014delegation}.
The established connection state is then handed to the constrained device and it can thus benefit from the high level of security.
Overall, if the given communication scenario allows for optimizing cryptography, especially resource constraints of devices and networks can be addressed. 
However, since this might impact interoperability, it should be further examined how those can be integrated into established protocols.

\subsection{Device Security} \label{sec:deviceSecurity}

Besides securing the network and communication, also individual devices need to be secured (cf.\ Section \ref{sec:deviceSec}).
In particular, the potentially easier physical access to devices in critical infrastructure \C{A3} poses a significant threat.
The device security must be considered on various levels to ensure trust in the system.
This includes hardware and software security as well as a proper root of trust.

\highlight{Hardware Security Mechanisms} are inevitable for trust into systems that are deployed in hostile environments to protect against manipulation of systems through direct physical access \C{A3}.
One prominent approach are secure elements \cite{schlapfer2019security}, which implement security-related mechanisms such as tamper-proof memory to store keys and certificates as well as secure random number generation in hardware.
Furthermore, secure elements can also improve performance through hardware acceleration and thus relieve resource-constrained devices \C{C4}.
Most notably, such optimizations concern rather resource-intensive algorithms such as those used in \ac{TLS}, making it easier to meet the legal requirement \C{R1} to use this protocol.  
However, it is important to consider that embedded secure elements can reduce updatability as only current security functions are implemented in hardware.
Further research in the area of hardware security mechanisms could improve on yet unsolved aspects such as the detection and prevention of hardware trojans and the protection against intellectual property theft through reverse engineering \cite{roy2023device}.

\highlight{Authenticated Code and Secure Boot} are further fundamental building blocks for the security of the overall system.
They are used to protect the software integrity of systems and thus, by extension, assure the authenticity \C{S3} of any interaction with that system and mitigate its manipulation through remote or direct physical interaction \C{A3}. 
For this, any code is authenticated via digital signatures or similar, which are then used during the boot process of the system to ensure that only authenticated code is executed \cite{frazelle2020securing}. 
However, a recent paper on secure boot in an automotive setting outlines various challenges that still need to be addressed \cite{sanwald2020secure}. 
In particular, an unbroken and properly rooted chain of trust throughout the entire boot process is of utmost importance.

\highlight{A Root of Trust} is inevitable for authentication \C{S3} across various levels and to enable further secure communication with the system.
When relying on root credentials, challenges during deployment \C{A1} and exclusion \C{A2} of devices have to be considered. 
Certificates, as extensively used in \ac{TLS} \cite{rfcTLS13}, cryptographically link an identity to a public key, which can then be used for authentication and key agreement, but revocation is non-trivial.
\ac{SIM} and \ac{eSIM} cards, are commonly used for authentication in mobile networks, but several security issues arise from a lack of authentication and protection of the communication between \ac{SIM} and modem \cite{zhao2021securesim}.
\ac{PUF} promise to be an invaluable tool to securely bind keys to a device.
Here, developments in the area of nanotechnology-based \ac{PUF}s and asymmetric keys derived from \ac{PUF}s are particularly interesting \cite{gao2020physical}.
As \ac{PUF}s might also provide tamper-resistance, such mechanisms further address the challenge of easier access to the system \C{A3}.
Moreover, recent open-source projects such as OpenTitan \cite{opentitan}, promise to close the remaining gaps between hardware and software security.
Here, updatability \C{R2} is also considered on the lowest layers and built into the architecture \cite{frazelle2020securing}.

\subsection{Network Knowledge} \label{sec:networkKnowledge}

Finally, detailed knowledge about a wireless network, especially w.r.t.\ topology and capabilities, provides ample opportunities to strengthen its secure and reliable use.
Beneficial knowledge ranges from general propagation conditions over context awareness of devices to redundancy considerations for applications.

\highlight{Propagation Conditions} lead to an irregular availability \C{S2} of a wireless network.
In addition, reaching a desired throughput to tackle network limitations \C{C3} requires sufficient signal power. 
Therefore, detailed network planning is mandatory for secure wireless communication.
For public networks, this planning process is conducted by an \ac{MNO} and the users are confronted with the results.
For shared and private networks, the critical infrastructure operator can optimize the receptions based on its applications and needs.
Conducting such an optimization is challenging and requires knowledge about the expected reception at different locations or areas.
Propagation models are well researched and can assist in this task~\cite{rappaport2017overview}.
However, conducting a targeted measurement campaign is often desirable~\cite{rademacher2021path}.
The resulting data (from models or campaigns) can be used to tackle the challenges of availability \C{S2} and mobility \C{C2}.

\highlight{Context Awareness} of a device in a wireless network plays a fundamental role in adjusting security mechanisms to current requirements and conditions.
The most prominent context is the position (geographical coordinates) of a device, which is particularly relevant when considering mobility \C{C2}.
Coordinates can be obtained either manually during roll-out (for stationary devices), automatically via technologies such as \ac{GPS}, or even with the communication technology itself~\cite{9665420}.
For 6G, the combination of communication and sensing is regarded as one major innovation~\cite{liu2022integrated}.
Other relevant context informations are the criticality of the application the device is used for, if the device has been recently moved, or even other environmental data such as the temperature.
In particular, such context information is useful to detect manipulation or unwanted movement of the devices, which addresses the challenge of easier physical access to devices \C{A3}. 
In addition, context information provides the opportunity to dynamically influence security parameters such as cryptographic algorithms based on the current context~\cite{hellaoui2020energy} to adjust to varying network limitations \C{C3} and available system resources \C{C4}. 

\highlight{Roaming and Redundancy} through the utilization of multiple wireless network technologies can be highly beneficial for critical infrastructures to improve reliability \C{C1} and availability \C{S2} in case of network failures or targeted attacks such as jamming.
To this end, a self-configuring multi-access gateway can be employed to meet the strict requirements in critical infrastructure \cite{mogensen2019implementation}.
In general, there are two possible scenarios for such a gateway.
First, multiple technologies are used simultaneously, which is particularly beneficial for mobility \C{C2} but also enhances confidentiality \C{S1}.
However, a mechanism is needed which distributes communication over the different technologies which increases the system complexity~\cite{8746482}.
When this mechanism is aware of the application context, delay-critical applications may use technologies such as 5G while low-priority data is transferred via \ac{LoRaWAN}.
Second, one technology is preferred and the others function purely as backup, thereby increasing the availability \C{S2} without additional system complexity. %

\section{Discussion and Conclusion} \label{sec:conclusion}

With the increasing trend towards wireless communication in general, a similar tendency and corresponding benefits can be observed for critical infrastructure. 
Most importantly, wireless communication affords the cost-effective interconnection of a multitude of devices without the need for rigid infrastructure. 
Moreover, it enables the mobility of devices and their deployment in remote areas. 
However, the diverse set of available wireless technologies and their different characteristics challenge the application of wireless communication in critical infrastructure.
Most importantly, easy access to the transmission medium and the use of shared or public infrastructure give rise to serious security challenges. 

In this paper, we categorized the relevant challenges that need to be addressed to enable the use of wireless communication in critical infrastructure alongside various dimensions.
Network and device constraints arise from characteristics of wireless communication as well as limitations of the network and the utilized devices. 
Moreover, network security challenges must be addressed with a focus on (end-to-end) confidentiality and integrity protection,  particularly critical availability, and authentication towards the network and communication peers. 
Novel device security challenges arise especially through the easier physical access to devices but also extend to the deployability and exclusion of devices. 
Lastly, application-specific requirements cover the need for flexibility of the system w.r.t.\ interoperability, updatability, and the consideration of future requirements to which a too-inflexible system cannot adapt.
This includes the stringent legal and regulatory requirements for critical infrastructure which could further tighten over time. 

Current approaches and research trends already provide a multitude of opportunities to secure the wireless communication of critical infrastructure by addressing the identified challenges. 
Network security mechanisms address reliability and limitations to some extent, but confidentiality and integrity protection are limited to hop-by-hop security.
For communication security on a higher layer, several approaches promise to enhance efficiency and address network and device limitations through beneficial configuration, extension, or optimization.
Within the scope of device security, several hardware and software security mechanisms can be identified as promising building blocks for the overall security. 
By combining them in a secure architecture with trust rooted in hardware, an unbroken chain of trust could be achieved. 
Moreover, detailed network knowledge is particularly promising for addressing challenges such as reliability and mobility.

Our work highlights the opportunities for securing wireless communication in critical infrastructure and identifies avenues for further research to fully capitalize on the benefits of wireless communication in critical infrastructure while guaranteeing strong security.
Overall, no single approach is sufficient to overcome all the challenges nor is this necessary.
Rather, an interplay of multiple solutions must be strived for to achieve a resilient and secure system. 
To this end, this paper aims to spark and guide further research efforts toward coherent yet adaptable solutions that address the unique challenges of securing wireless communication in critical infrastructure.

\highlight{Acknowledgments}
This work has been funded by the German Federal Office for Information Security (\ac{BSI}) under project funding reference numbers 01MO23003A, 01MO23003B, 01MO23003C and 01MO23003D (PlusMoSmart).
The responsibility for the content of this publication lies with the authors.

\begin{acronym}[]
	\acro{3GPP}{3rd Generation Partnership Project}
	\acro{AC}{Access Category}
	\acro{ACK}{acknowledgement}
	\acro{AI}{Abstract Interface}
	\acro{AIFS}{Arbitration Interframe Space}
	\acro{AODV}{Ad hoc On-Demand Distance Vector Routing Protocol}
	\acro{AP}{Access Point}
	\acro{API}{application programming interface}
	\acro{APN}{Access Point Name}
	\acro{ARP}{Address Resolution Protocol}
	\acro{ARQ}{Automatic Repeat reQuest}
	\acro{AS}{Autonomous System}
	\acro{ASCII}{American Standard Code for Information Interchange}
	\acro{ATIS}{Alliance for Telecommunications Industry Solutions}
	\acro{ATM}{Asynchronous Transfer Mode}
	\acro{AMR}{Automatic Meter Reading}
	\acro{B.A.T.M.A.N.}{Better Approach to MANET}
	\acro{BER}{Bit Error Rate}
	\acro{BFS-CA}{Breadth First Search Channel Assignment}
	\acro{BGP}{Border Gateway Protocol}
	\acro{BPSK}{Binary Phase-Shift Keying}
	\acro{BRA}{Bidrectional Routing Abstraction}
	\acro{BS}{Base station}
	\acro{BSI}{Bundesamt für Sicherheit in der Informationstechnik}
	\acro{BSSID}{Basic Service Set Identification}
	\acro{BTS}{Base Transceiver Station}
	\acro{CA}{Channel Assignment}
	\acro{CAPEX}{capital expenditure}
	\acro{CAPWAP}{Control And Provisioning of Wireless Access Points}
	\acro{CARD}{Channel Assignment with Route Discovery}
	\acro{CAS}{Channel Assignment Server}
	\acro{CCA}{Clear Channel Assessment}
	\acro{CDMA}{Code Division Multiple Access}
	\acro{CF}{CompactFlash}
	\acro{CIDR}{Classless Inter-Domain Routing}
	\acro{CLICA}{Connected Low Interference Channel Assignment}
	\acro{CLI}{Command Line Interface}
	\acro{COTS}{Commercial Off-the-Shelf}
	\acro{CPE}{Customer Premises Equipment}
	\acro{CPU}{Central Processing Unit}
	\acro{CRAHN}{Cognitive Radio Ad-Hoc Network}
	\acro{CRCN}{Cognitive Radio Cellular Network}
	\acro{CR}{Cognitive Radio}
	\acro{CR-LDP}{Constraint-based Routing Label Distribution Protocol}
	\acro{CRN}{Cognitive Radio Network}
	\acro{CRSN}{Cognitive Radio Sensor Network}
	\acro{CRVN}{Cognitive Radio Vehicular Network}
	\acro{CSMA/CA}{Carrier Sense Multiple Access/Collision Avoidance}
	\acro{CSMA}{Carrier Sense Multiple Access}
	\acro{CSMA/CD}{Carrier Sense Multiple Access/Collision Detection}
	\acro{CSV}{Comma-Separated Values}
	\acro{CTA}{Centralized Tabu-based Algorithm}
	\acro{CW}{Contention Window}
	\acro{CWLAN}{Cognitive Wireless Local Area Network}
	\acro{CWMN}{Cognitive Wireless Mesh Network}
	\acro{DAD}{Duplicate Address Detection}
	\acro{DCF}{Distributed Coordination Function}
	\acro{DCiE}{Data Center Infrastructure Efficiency}
	\acro{DDS}{Direct digital synthesizer}
	\acro{DFS}{Dynamic Frequency Selection}
	\acro{DGA}{Distributed Greedy Algorithm}
	\acro{DHCP}{Dynamic Host Configuration Protocol}
	\acro{DIFS}{Distributed Interframe Space}
	\acro{DMesh}{Directional Mesh}
	\acro{D-MICA}{Distributed Minimum Interference Channel Assignment}
	\acro{DR}{Designated Router}
	\acro{DSA}{Dynamic Spectrum Allocation}
	\acro{DSLAM}{Digital Subscriber Line Access Multiplexer}
	\acro{DSL}{Digital Subscriber Line}
	\acro{DSR}{Dynamic Source Routing Protocol}
	\acro{DSSS}{Direct-Sequence Spread Spectrum}
	\acro{DTCP}{Dynamic Tunnel Configuration Protocol}
	\acro{DVB}{Digital Video Broadcast}
	\acro{DVB-H}{Digital Video Broadcast - Handheld}
	\acro{DVB-RCS}{Digital Video Broadcast - Return Channel Satellite}
	\acro{DVB-S2}{Digital Video Broadcast - Satellite - Second Generation}
	\acro{DVB-S}{Digital Video Broadcast - Satellite}
	\acro{DVB-SH}{Digital Video Broadcast - Satellite services to Handhelds}
	\acro{DVB-T2}{Digital Video Broadcast - Second Generation Terrestrial}
	\acro{DVB-T}{Digital Video Broadcast - Terrestrial}
	\acro{E2CARA-TD}{Energy Efficient Channel Assignment and Routing Algorithm – Traffic Demands}
	\acro{ECN}{Explicit Congestion Notification}
	\acro{ECDF}{Empirical Cumulative Distribution Function}
	\acro{EDCA}{Enhanced Distributed Coordination Access}
	\acro{EDCF}{Enhanced Distributed Coordination Function}
	\acro{EGP}{Exterior Gateway Protocol}
	\acro{EICA}{External Interference-Aware Channel Assignment}
	\acro{EIFS}{Extended Interframe Space}
	\acro{EIGRP}{Enhanced Interior Gateway Routing Protocol}
	\acro{EIRP}{Equivalent Isotropically Radiated Power}
	\acro{EPI}{energy proportionality index}
	\acro{ERO}{Explicit Route Object}
	\acro{ETSI}{European Telecommunications Standards Institute}
	\acro{ETT}{Expected Transmission Time}
	\acro{ETX}{Expected Transmission Counts}
	\acro{EUI}{Extended Unique Identifier}
	\acro{FCC}{Federal Communications Commission}
	\acro{FCS}{Frame Check Sequence}
	\acro{FDD}{Frequency Division Duplex}
	\acro{FDMA}{Frequency Division Multiple Access}
	\acro{FEC}{Forward Error Correction}
	\acro{FIFO}{First-In-First-Out}
	\acro{FLOPS}{Floating-Point Operations Per Second}
	\acro{FRR}{Fast Reroute}
	\acro{FSL}{Free-Space Loss}
	\acro{FSPL}{Free-Space Path Loss}
	\acro{GAN}{Generic Access Network}
	\acro{GDP}{Gross Domestic Product}
	\acro{GEO}{Geosynchronous Earth Orbit}
	\acro{GMPLS}{Generalized Multiprotocol Label Switching}
	\acro{GPS}{Global Positioning System}
	\acro{GRE}{Generic Routing Encapsulation}
	\acro{GSE}{Generic Stream Encapsulation}
	\acro{GSM}{Global System for Mobile Communications}
	\acro{GW}{Gateway}
	\acro{HAP}{High-Altitude Platform}
	\acro{HCCA}{HCF controlled channel access}
	\acro{HCF}{Hybrid Coordination Function}
	\acro{HLR}{Home Location Register}
	\acro{HOL}{Head-of-line}
	\acro{HOLSR}{Hieracical Optimised Link State Routing}
	\acro{HPC}{hardware performance counters}
	\acro{HSLS}{Hazy-Sighted Link State Routing Protocol}
	\acro{HWMP}{Hybrid Wireless Mesh Protocol}
	\acro{IAX2}{Inter-Asterisk eXchange Version 2}
	\acro{IBSS}{Independent Basic Service Set}
	\acro{ICMP}{Internet Control Message Protocol}
	\acro{ICT}{Information and Communication Technologie}
	\acro{IEEE}{Institute of Electrical and Electronics Engineers}
	\acro{IE}{Information Element}
	\acro{IETF}{Internet Engineering Task Force}
	\acro{IETF}{The Internet Engineering Task Force}
	\acro{IFS}{Interframe Space}
	\acro{ITU}{International Telecommunication Union}
	\acro{IGP}{Interior Gateway Protocol}
	\acro{IGRP}{Interior Gateway Routing Protocol}
	\acro{ILP}{Integer Linear Programming}
	\acro{ILS}{Iterated Local Search}
	\acro{IPFIX}{IP Flow Information Export}
	\acro{IP}{Internet Protocol}
	\acro{IPv4}{Internet Protocol}
	\acro{IPv6}{Internet Protocol, Version 6}
	\acro{ISI}{Inter-symbol interference}
	\acro{IS-IS}{Intermediate system to intermediate system}
	\acro{ISM}{Industrial, Scientific and Medical}
	\acro{ISP}{Internet Service Provider}
	\acro{JSON}{JavaScript Object Notation}
	\acro{KPI}{Key-Performance-Indicator}
	\acro{LAA}{Licensed-Assisted Access}
	\acro{LDC}{Least Developed Countries}
	\acro{LA-CA}{Load-Aware Channel Assignment}
	\acro{LCOS}{LANCOM Operating System}
	\acro{LDP}{Label Distribution Protocol}
	\acro{Ld}{Log-distance}
	\acro{LDPL}{Log-distance path loss}
	\acro{LEO}{Low Earth Orbit}
	\acro{LER}{Label Edge Router}
	\acro{LGI}{Long Guard Interval}
	\acro{LLTM}{Link Layer Tunneling Mechanism}
	\acro{LMA}{Local Mobility Anchor}
	\acro{LMP}{Link Management Protocol}
	\acro{LoS}{Line of Sight}
	\acro{LOS}{Line of Sight}
	\acro{LQF}{Longest-Queue-First}
	\acro{LS}{Link State}
	\acro{LSP}{Label-Switched Path}
	\acro{LSR}{Label-Switched Router}
	\acro{LST}{Link-State-Table}
	\acro{LTE}{Long Term Evolution}
	\acro{LTE-M}{\ac{LTE} Machine Type Communication}
	\acro{LWAPP}{Lightweight Access Point Protocol}
	\acro{MAC}{Media Access Control}
	\acro{MAG}{Mobile Access Gateway}
	\acro{MANET}{Mobile Adhoc Network}
	\acro{MBMS}{Multimedia Broadcast Multicast Service}
	\acro{MCG}{multi-conflict graph}
	\acro{MCI-CA}{Matroid Cardinality Intersection Channel Assignment}
	\acro{MCS}{Modulation and Coding Scheme}
	\acro{MDR}{MANET Designated Router}
	\acro{MEO}{Medium Earth Orbit}
	\acro{MICS}{Media Independent Command Service}
	\acro{MIES}{Media Independent Event Service}
	\acro{MIHF}{Media Independent Handover Function}
	\acro{MIHF++}{Media Independent Handover Function++}
	\acro{MIH}{Media Independent Handover}
	\acro{MIIS}{Media Independent Information Service}
	\acro{MILP}{Mixed Integer Linear Programming}
	\acro{MIMF}{Media Independent Messaging Function}
	\acro{MIMO}{Multiple Input Multiple Output}
	\acro{MNO}{Mobile Network Operator}
	\acro{MIMS}{Media Independent Messaging Service}
	\acro{MIPS}{Million Instruction Per Second}
	\acro{MMF}{Mobility Management Function}
	\acro{MMS}{Manufacturing Message Specification}
	\acro{MN}{Mesh Node}
	\acro{MonF}{Monitoring Function}
	\acro{MPDU}{MAC Protocol Data Unit}
	\acro{MPEG}{Moving Picture Experts Group}
	\acro{MPE}{Multi Protocol Encapsulation}
	\acro{MPLCG}{Multi-Point Link Conflict Graph}
	\acro{MPLS}{Multiprotocol Label Switching}
	\acro{MPLS}{Multi Protocol Label Switching}
	\acro{MPLS-TE}{Multi Protocol Label Switching - Traffic Engineering}
	\acro{MP}{Merge Point}
	\acro{MPR}{Multipoint Relay}
	\acro{MR-MC WMN}{Multi-Radio Multi-Channel Wireless Mesh Network}
	\acro{MSC}{Mobile-services Switching Centre}
	\acro{MSDU}{MAC Service Data Unit}
	\acro{MSTP}{Mobility Services Transport Protocol}
	\acro{MT}{Mobile Terminal}
	\acro{MTU}{Maximum Transmission Unit}
	\acro{MQTT}{MQ Telemetry Transport}
	\acro{NAV}{Network Allocation Vector}
	\acro{ns-3}{network simulator 3}
	\acro{NB-IoT}{Narrowband \ac{IoT}}
	\acro{NBMA}{Non-broadcast Multiple Access}
	\acro{NetEMU}{Network Emulator}
	\acro{NIDD}{Non-\ac{IP} Data Delivery}
	\acro{NLOS}{None Line of Sight}
	\acro{NMEA}{National Marine Electronics Association}
	\acro{NPC}{Normalized Power Consumption}
	\acro{NP}{Nondeterministic Polynomial Time}
	\acro{NSIS}{Next Steps in Signaling}
	\acro{NTP}{Network Time Protocol Unit}
	\acro{OFDMA}{Orthogonal Frequency Division Multiple Access}
	\acro{OFDM}{Orthogonal Frequency Division Multiplex}
	\acro{OLSR}{Optimized Link State Routing}
	\acro{OPEX}{operational expenditure}
	\acro{OSA}{Opportunistic Spectrum Access}
	\acro{OSI}{Open Systems Interconnection}
	\acro{OSPF}{Open Shortest Path First}
	\acro{OSPF-TE}{Open Shortest Path First - Traffic Engineering}
	\acro{OVS}{Open vSwitch}
	\acro{P2MP}{Point To Multipoint}
	\acro{P2P}{Point To Point}
	\acro{PA}{Power amplifier}
	\acro{PCE}{Path Computation Element}
	\acro{PCEP}{Path Computation Element Protocol}
	\acro{PCF}{Path Computation Function}
	\acro{PCF}{Point Coordination Function}
	\acro{PDR}{Packet Delivery Ratio}
	\acro{PDV}{Packet Delay Variation}
	\acro{PER}{Packet Error Rate}
	\acro{PLCP}{Physical Layer Convergence Protocol}
	\acro{PLL}{Phase-Locked Loop}
	\acro{PL}{path loss}
	\acro{PLR}{Point of Local Repair}
	\acro{PMIP}{Proxy Mobile IP}
	\acro{PoE}{Power over Ethernet}
	\acro{PPDU}{Physical Protocol Data Unit}
	\acro{PPP}{Point-to-Point Protocol}
	\acro{PSTN}{Public Switched Telephone Network}
	\acro{PTP}{Precision Time Protocol}
	\acro{PUE}{Power Usage Effectiveness}
	\acro{PU}{Primary User}
	\acro{QAM}{Quadrature amplitude modulation}
	\acro{QoS}{Quality of Service}
	\acro{RAN}{Radio Access Network}
	\acro{RAND}{Random Channel Assignment}
	\acro{RERR}{Route Error}
	\acro{RFC}{Request for Comments}
	\acro{RIPng}{Routing Information Protocol next generation}
	\acro{RIP}{Routing Information Protocol}
	\acro{RPC}{Remote Procedure Call}
	\acro{RPP}{Received Packet Power}
	\acro{RREP}{Route Reply}
	\acro{RREQ}{Route Request}
	\acro{RSSI}{Received Signal Strength Indication}
	\acro{RSS}{Received Signal Strength}
	\acro{RSVP}{Resource ReSerVation Protocol}
	\acro{RSVP-TE}{Resource ReSerVation Protocol - Traffic Engineering}
	\acro{RTK}{Real Time Kinematic}
	\acro{RTP}{Real-time Transport Protocol}
	\acro{RTS}{Ready-To-Send}
	\acro{RTT}{Round Trip Time}
	\acro{RRC}{Radio Resource Control}
	\acro{SAA}{Stateless Address Autoconfiguration}
	\acro{SAPOS}{Satellitenpositionierungsdienst der deutschen Landesvermessung}
	\acro{SAP}{Service Access Point}
	\acro{SBC}{Single-Board Computer}
	\acro{SBM}{Subnetwork Bandwidth Manager}
	\acro{SBR}{System zur Bestimmung des Richtungsfehlers}
	\acro{SCADA}{Supervisory Control and Data Acquisition}
	\acro{SC-FDMA}{Single Carrier Frequency Division Multiple Access}
	\acro{SDMA}{Space-division multiple access}
	\acro{SDN}{Software Defined Networking}
	\acro{SDR}{Software Defined Radio}
	\acro{SDWN}{Software Defined Wireless Networks}
	\acro{SENF}{Simple and Extensible Network Framework}
	\acro{SGI}{Short Guard Intervall}
	\acro{SIFS}{Short Interframe Space}
	\acro{SINR}{Signal-to-Noise-plus-Interference Ratio}
	\acro{SIP}{Session Initiation Protocol}
	\acro{SISO}{Single Input Single Output}
	\acro{SNR}{signal-to-noise ratio}
	\acro{SONET}{Synchronous Optical Networking}
	\acro{SPoF}{Single Point of Failure}
	\acro{SSID}{Service Set Identifier}
	\acro{STP}{Spanning Tree Protocol}
	\acro{SU}{Secondary User}
	\acro{TCP}{Transmission Control Protocol}
	\acro{TPC}{Transmission Power Control}
	\acro{TC}{Topology Control}
	\acro{TDMA}{Time Division Multiple Access}
	\acro{TEEER}{Telecommunications Equipment Energy Efficiency Rating}
	\acro{TEER}{Telecommunications Energy Efficiency Ratio}
	\acro{TETRA}{Terrestrial Trunked Radio}
	\acro{TE}{Traffic Engineering}
	\acro{TIM}{Technology Independend Monitoring}
	\acro{TLV}{Type-Length-Value}
	\acro{TLS}{Transport Layer Security}
	\acro{TORA}{Temporally-Ordered Routing Algorithm}
	\acro{ToS}{Type of Service}
	\acro{TSFT}{Time Synchronization Function Timer}
	\acro{TTL}{Time to live}
	\acro{TVWS}{TV White Space}
	\acro{TXOP}{Transmit opportunity}
	\acro{UAV}{Unmanned Aerial Vehicle}
	\acro{UDLR}{Unidirectional Link Routing}
	\acro{UDL}{Unidirectional Link}
	\acro{UDP}{User Datagram Protocol}
	\acro{UDT}{Unidirectional Technology}
	\acro{UE}{User Equipment}
	\acro{UHF}{Ultra High Frequency}
	\acro{UMA}{Unlicensed Mobile Access}
	\acro{UMTS}{Universal Mobile Telecommunications System}
	\acro{U-NII}{Unlicensed National Information Infrastructure}
	\acro{UPS}{Uninterruptible Power Supply}
	\acro{UP}{User Priorities}
	\acro{USB}{Univeral Serial Bus}
	\acro{USO}{Universal Service Obligation}
	\acro{USRP}{Universal Software Radio Peripheral}
	\acro{VCO}{Voltage-Controlled Oscillator}
	\acro{VoIP}{Voice-over-IP}
	\acro{VPN}{Virtual Private Network}
	\acro{WAN}{Wide Area Network}
	\acro{WBN}{Coordinated Wireless Backhaul Network}
	\acro{WDS}{Wireless Distribution System}
	\acro{WiBACK}{Wireless Back-Haul}
	\acro{Wi-Fi}{Wireless Fidelity}
	\acro{WiLD}{WiFi-based Long Distance}
	\acro{WiMAX}{Worldwide Interoperability for Microwave Access}
	\acro{WISPA}{Wireless Internet Service Provider Assocication}
	\acro{WISP}{Wireless Internet Service Provider}
	\acro{WLAN}{Wireless Local Area Network}
	\acro{WLC}{Wireless LAN Controller}
	\acro{WMN}{Wireless Mesh Network}
	\acro{WMN}{Wireless Mesh Network}
	\acro{wmSDN}{Wireless Mesh Software Defined Network}
	\acro{WNIC}{Wireless Network Interface Controller}
	\acro{WN}{WiBACK Node}
	\acro{WRAN}{Wireless Regional Area Network}
	\acro{WSN}{Wireless Sensor Network}
	\acro{ZigBee}{ZigBee Alliance IEEE 802.15.4}
	\acro{ZPR}{Zone Routing Protocol}
	\acro{ABP}{Activation by Personalization}
	\acro{ADR}{Adaptive Data Rate}
	\acro{IoT}{Internet of Things}
	\acro{LPWAN}{low-power wide-area network}
	\acro{LoRa}{Long Range}
	\acro{LoRaWAN}{Long Range Wide Area Network}
	\acro{GPRS}{General Packet Radio Service}
	\acro{CEP}{Circular Error Probable}
	\acro{EIRP}{equivalent isotropically radiated power}
	\acro{ITM}{Longley-Rice Irregular Terrain Model}
	\acro{SF}{Spreading Factor}
	\acro{ASDU}{Application Service Data Unit}
	\acro{IEC 104}{IEC 60870-5-104}
	\acro{SIM}{Subscriber Identity Module}
	\acro{eSIM}{embedded \ac{SIM}}
	\acro{PKI}{Public Key Infrastructure}
	\acro{FHSS}{Frequency-Hopping Spread Spectrum}
	\acro{UWB}{Ultra-Wideband}
	\acro{PUF}{Physical Unclonable Functions}
\end{acronym}

\end{document}